\documentclass[reprint,prb,aps,twocolumn,superscriptaddress, 10pt]{revtex4-2}
\usepackage{amsmath}
\usepackage[english]{babel}
\usepackage{graphicx}
\usepackage{color}
\usepackage{sidecap}
\usepackage[dvipsnames]{xcolor}

\usepackage[colorlinks,
            citecolor = blue, 
						linkcolor = blue,
						urlcolor  = blue,
						anchorcolor = blue,
						breaklinks = true
						]{hyperref}

\usepackage{cleveref}

\usepackage{soul}
\usepackage{ amssymb }
\newif\ifdraft
\drafttrue
\def \Marylanda{Department of Materials Science and Engineering, University of Maryland, College Park, MD 20742, USA}
\def \Basel{Department of Physics, University of Basel, Klingelbergstrasse 82, Basel CH-4056, Switzerland}
\def \WM{Department of Physics, University of Wisconsin-Madison, USA}

\begin{document}

% ----- Title -----

\title{Electronic crystals in layered materials}

\author{Y. Zhou}
\affiliation{\Marylanda} %, \Marylandb}

\author{I. Esterlis}
\affiliation{\WM}

\author{T. Smole\'nski}
\affiliation{\Basel}

\begin{abstract} 
In modern two-dimensional (2D) materials, such as graphene-based systems and atomically-thin transition-metal dichalcogenides, the interplay of strong electronic correlations, tunable moiré superlattices, and nontrivial band topology has given rise to rich phase diagrams and collective phenomena. Among the novel phases that have been realized, electronic crystals -- states of matter in which itinerant electrons spontaneously crystallize -- play a particularly prominent role. In this Review, we summarize the current status of electron crystallization in van der Waals heterostructures, with emphasis on the experimental platforms and measurement techniques that enable their study. We also highlight open questions and outline future directions that may elucidate more of the fascinating properties of electronic crystals. 
\end{abstract}

\maketitle

%Things to include (work in progress):
%
%\begin{enumerate}
%\textbf{Fundamentals of electronic crystallization}
%{\color{blue}
%\begin{itemize}
%\item Why electrons crystallize? What is the crystal? What is the phase diagram? Role of intermediate phases.
%\item Magnetism of Wigner crystal and of intermediate phases; Pomeranchuk effect.
%\item Difference between electronic crystalization in the presence/absence of external moir\'e potential: from Wigner crystals, through generalized Wigner crystals, to Mott states (or charge-transfer insulators).
%\item Role of effective mass/screening: here we can discuss experiments performed in MoSe$_2$ bilayer for holes, which have a few times larger effective mass, giving rise to much larger critical electron density for the Wigner crystal.
%\end{itemize}
%}

%{\color{red}IE: Should highlight important open questions that can/should/will be addressed in the field, both in electron crystallization and beyond.}

%{\color{violet}{TS: can we add $4\pi\epsilon_0\epsilon_r$ and proper units to the formulas in Sec. IIA (I feel that it is always super useful for experimentalists)}}

\section{Introduction}

\begin{figure*}[t]
	\includegraphics[width=0.99\textwidth]{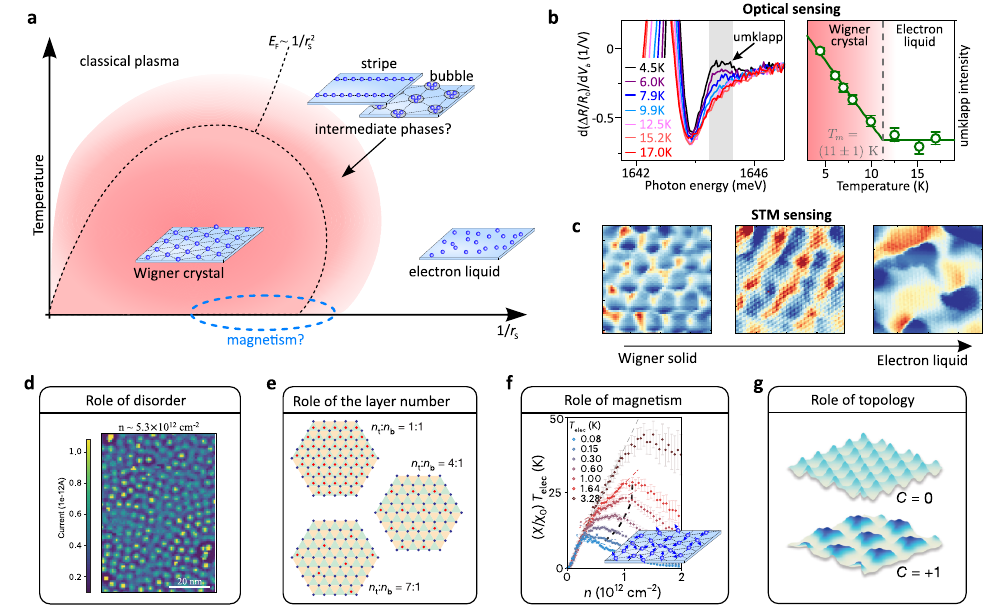}
	\caption{{\bf Properties and sensing of an electronic phase diagram} ({\bf a}) Schematic phase diagram of the idealized 2DEG as a function of $1/r_s$ (tuned by varying the electron density) and temperature, highlighting open questions. ({\bf b}) Optical detection of an umklapp scattering peak in the reflection contrast measurements on monolayer MoSe$_2$, indicating the formation of a WC; adapted from Ref.~\cite{smolenski2021}. ({\bf c}) STM imaging of the WC and stripe phases in bilayer graphene system in a X T magnetic field; adapted from Ref.~\cite{tsui2024}. ({\bf d}) - ({\bf g}) New features of electron crystal formation that can be probed with 2D materials: ({\bf d}) disorder effects captured by STM imaging on bilayer MoSe$_2$, adapted from Ref.~\cite{xiang2025}; multilayer crystals, as observed in bilayer MoSe$_2$ \cite{zhou2021}, spin/valley magnetism, recently studied in monolayer MoSe$_2$ in Ref.~\cite{sung2025}; and the interplay between electron crystallization and band topology, as studied in a moiré pentalayer graphene system \cite{su2025}.} 
    \label{fig:Fig1}
\end{figure*}

At sufficiently low carrier densities, a homogeneous electron system can spontaneously ``freeze" into an ordered electronic ``Wigner" crystal (WC), driven by the dominance of electron-electron interactions over kinetic energy. First predicted by Eugene Wigner nearly a century ago \cite{wigner1934}, these charge-ordered states are a paradigm of strongly correlated electronic states of matter (Fig.~\ref{fig:Fig1}{\bf a}). 

While the possibility of electronic crystallization in metallic systems was a theoretical curiosity in Wigner's time, it has become clear that electron crystals play a prominent role in the phase diagrams of modern two-dimensional (2D) quantum materials based on van der Waals (vdW) heterostructures, including graphene-based systems and atomically-thin transition-metal dichalcogenides (TMDs). Experimental realizations of electron crystals in 2D materials to date include single \cite{smolenski2021,sung2025} and bilayer \cite{zhou2021,xiang2025} TMDs, twisted TMDs with moiré superlattices \cite{regan2020,xu2020,huang2021,jin2021,li2021,li2021b,li2024,shabani2021}, crystalline graphene multilayers \cite{tsui2024} and moiré graphene \cite{walters2025, su2025}. These remarkable discoveries, enabled by advances in device fabrication and novel sensing and detection techniques (Fig.~\ref{fig:Fig1}{\bf b}, {\bf c}), have led to a resurgence of interest in electron crystals. Moreover, the  prominence of these phases suggests that their detailed characterization is crucial for a comprehensive understanding of the phase diagrams of charge-tunable vdW devices. 
These systems are also especially promising for addressing fundamental questions associated with electron crystallization, such as the effects of quenched disorder, their spin and valley magnetism, crystallization in multilayer systems, and interplay with band topology (see Figs.~\ref{fig:Fig1}{\bf d}-{\bf g}).

%--  not only as interesting states of matter in their own right but also because they often appear in close proximity to other fascinating phases. A detailed characterization of these electron crystal phases therefore appears crucial for a comprehensive understanding of the phase diagrams of 2D materials.

The purpose of this article is to review recent advances in electron crystallization in layered 2D materials.  Section \ref{sec:wc_general} summarizes the fundamental features of 2D WCs and explains why modern 2D platforms have proved especially fruitful for their study. In Section \ref{sec:wc_sensing} we describe the experimental detection methods special to 2D materials that have been especially important in the discovery and study of electron crystals. 
Finally, in Section \ref{sec:wc_outlook} we conclude with a discussion of more exotic types of electron crystals that are now the subject of active theoretical and experimental investigation. 

%The study of electron crystallization has a long history. %, and many excellent reviews exist in the literature \cite{}.  
%Here, our focus is on the most recent developments in 2D materials, with an emphasis on experimental systems and capabilities, as well as perspectives for future directions. 

\section{WCs in layered materials}
\label{sec:wc_general}

\subsection{WCs in an ideal setting}

%\textcolor{blue}{Need a introductory figure depciting the idea and key concepts (i.e., theoretical phase diagram, etc.)}

In the simplest situation, the low-carrier density state of a 2D semiconductor in the effective mass approximation is described  by the homogeneous 2D electron gas (2DEG). The important energy scales are the Fermi energy $E_F = \hbar^2 / m^* a^2$ \footnote{We assume an isotropic, two-fold (spin or valley) degenerate electronic energy band.} and the Coulomb energy $V_C = e^2 / 4\pi \epsilon_0 \epsilon_r a$, where $m^*$ is the effective mass, $a$ is interparticle distance (defined in relation to the 2D electron density $n = 1/\pi a^2$), and $\epsilon_r$ is the dielectric constant of the environment in which the 2D layer is embedded. The single dimensionless parameter that determines the ground state properties of the system is $r_s \equiv V_C / E_F = a/ a_B^*$, where $a_B^* = 4\pi \epsilon_0 \epsilon_r \hbar^2 / m^* e^2$ is the effective Bohr radius of the semiconductor. Implicit in this description is a charge-neutralizing background, which is provided by the gate electrodes in the 2D materials under consideration. 

In the high-density limit $r_s \to 0$, the kinetic energy dominates over the interaction energy and the important physics is the metallic screening of the long-range Coulomb interaction, which is adequately captured within the random-phase approximation (RPA) \footnote{Another interesting possibility at high density is that of Kohn-Luttinger superconductivity \cite{chubukov2017}.}. In the opposite limit of low density, $r_s \to \infty$, the electron-electron interactions dominate over the kinetic energy and the electrons order into a triangular WC lattice to minimize their mutual Coulomb repulsion \cite{bonsall1977}. Quantum Monte Carlo (QMC) calculations place the transition between homogeneous liquid and WC phase at $r_s \approx 37$ \cite{tanatar1989,attaccalite2002,drummond2009,azadi2024,smith2024}, which is in good agreement with recent experiments \cite{smolenski2021,falson2022,sung2025}. 

However, even in this simplest 2DEG realization, there remain a number of important open questions concerning the ground state phase diagram. As $r_s \to \infty$, residual exchange couplings between the electrons localized on the WC lattice sites lead to a ferromagnetic ground state, as can be deduced from semiclassical considerations \cite{roger1984,katano2000,chakravarty1999,voelker2001}. However, as the system approaches the melting transition with decreasing $r_s$, larger multiparticle ring exchange processes become important, leading to highly frustrated and complex exchange dynamics \cite{drummond2009, smith2024, bernu2001, misguich1998}. Despite both theoretical and experimental progress, the question of the magnetic ground state of the WC near the melting point is not settled. Another central question concerns the role of intermediate phases between the WC and the homogeneous electron liquid. A direct first-order transition is forbidden due to the long-range Coulomb energy penalty in the associated two-phase coexistence region~\cite{spivak2004}, which has led to the proposal of intermediate mixed or micro-emulsion phases of electron liquid and crystal \cite{spivak2004,jamei2005,joy2023}, a proposal that has recently received experimental support \cite{sung2025}. The distinct possibility of an intermediate metallic electron crystal phase, arising as a self-doping instability of the WC, was also recently investigated \cite{kim2024}.  The situation is summarized in the schematic phase diagram in Fig.~\ref{fig:Fig1}{\bf a}. 

As will be elaborated upon below, the phase diagram is significantly enriched in 2D vdW materials, where additional axes --  including disorder, layer number, and band topology -- lead to new possibilities for electron crystallization. 

\subsection{2D materials platforms}

%\textcolor{blue}{Figure illustrating the key ingrediants responsible for enhnced stability of correlations in 2D (screening, masses).}

%\textcolor{blue}{Circle that depicts all systems we discuss (review style).}

\begin{table}[b]
\caption{\label{tab:params}
Representative material parameters for electrons ($e$) and holes in ($h$) in various TMDs, compared with a few quantum well systems. Here $m_*/m_e$ is the band mass in units of the free electron mass; $\epsilon_r$ is an effective dielectric constant of the semiconductor environment, the precise value of which depends on details of the heterostructure and, unless otherwise specified, we have put $\epsilon_r \approx 4.5$ as a representative value for different  TMDs \cite{mak2022}; $n_c$ is  the theoretically estimated critical density for WC formation corresponding to $r_s = 37$; $\mu$ is the highest reported mobility in the corresponding material.}
\begin{ruledtabular}
\begin{tabular}{ccccc}
 &$m_*/m_e$  &$\epsilon_r$ & $n_c$ ($10^{11} \text{ cm}^{-2}$) & $\mu$ (cm$^2$/Vs)\\
\hline
MoSe$_2$ (\textit e) & 0.7 \cite{sung2025} & 4.5  & 2.0 & 3,000 \cite{liu2025} \\
biMoSe$_2$ (\textit h)  & 1.26 \cite{xiang2025} & 2.6 \cite{xiang2025} & 20 & ? \\
WSe$_2$ (\textit h)  & 0.45 \cite{Pack2024} & 4.5 & 0.83 & 80,000 \cite{Pack2024} \\
WS$_2$ (\textit h)  & 0.35 \cite{tanabe2016} & 4.5 & 0.5 & 2,000 \cite{wang2021} \\
\hline
AlAs (\textit e) & 0.46 \cite{hossain2020} & 10 \cite{hossain2020} & 0.18  & 2.4$\times 10^6$ \cite{chung2018} \\
ZnO  (\textit e) & 0.3  \cite{falson2022} & 8.5   \cite{falson2022} & 0.10  & 6$\times 10^5$ \cite{falson2022} \\
GaAs (\textit e) & 0.067 \cite{hossain2020} & 13 \cite{hossain2020} & 0.002  & 57$\times 10^6$ \cite{chung2022_electron} \\
\end{tabular}
\end{ruledtabular}
%\footnotetext[1]{Ref.~\onlinecite{sung2025}}
%\footnotetext[2]{Ref.~\onlinecite{liu2025}}
%\footnotetext[3]{Ref.~\onlinecite{xiang2025}}
%\footnotetext[4]{Ref.~\onlinecite{Pack2024}}
%\footnotetext[5]{Ref.~\onlinecite{chung2022}}
\end{table}

\begin{figure}[t]
	\includegraphics[width=0.49\textwidth]{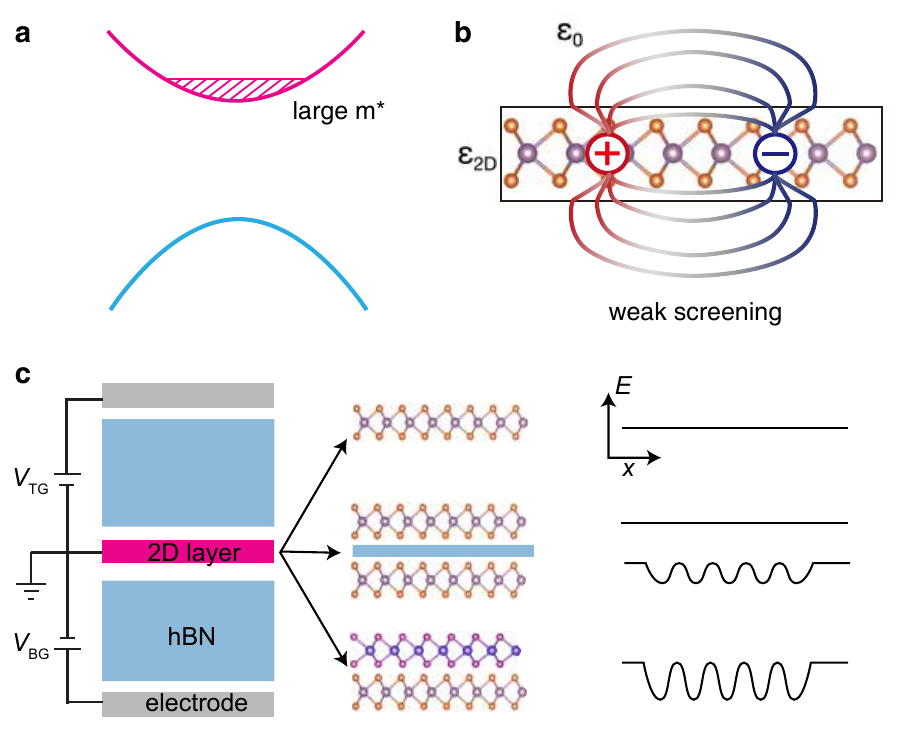}
	\caption{{\bf Properties of two-dimensional semiconductors} ({\bf a}) 2D materials can feature electronic bands with relatively large effective mass. ({\bf b}) The reduced dimensionality leads to strong Coulomb interactions among charged carriers. This, for instance, results in strong correlation effects and tightly bound excitons with large oscillator strengths. ({\bf c}) In a prototypical device structure, the 2D heterostructures are encapsulated under study in dielectric 2D layers such as hBN and gated with top and bottom electrodes, allowing independent control over carrier density and electric field. The heterostructure may consist of a monolayer, an electrostatically coupled bilayer, or bilayers in direct contact hosting a moiré superlattice. The corresponding electronic and/or excitonic potential is shown on the right of the panel. These configurations exhibit distinct electronic and excitonic properties: for instance, a monolayer preserves continuous symmetry, while a moiré superlattice breaks it into discrete symmetries. The bilayer system lies in between, capable of exhibiting either continuous or discrete symmetry depending on the specific configuration, such as interlayer coupling strength.} \label{fig:Fig2}
\end{figure}

Charge-tunable vdW heterostructures assembled by combining together individual layers of various 2D materials are especially well-suited to the study of WC phases. In the case of graphene-based systems, the highly adjustable electronic bands -- which can be significantly flattened using magnetic fields, displacement fields in the case of multilayers with rhombohedral stacking configuration~\cite{han2023,lu2024}, or by forming narrow bands in moir\'e systems -- lead to situations where interaction effects are dominant. In TMD systems, electronic crystallization can occur spontaneously even in the translationally-invariant (within the effective mass approximation) monolayer limit owing to relatively large effective masses and reduced dielectric screening, which combine to give relatively high critical densities for WC formation (see Fig.~\ref{fig:Fig2}{\bf a}, {\bf b} and Table \ref{tab:params}). %Its stability can be further enhanced by combining two TMD monolayers in a bilayer structure, which further enhances correlation effects, enabling reaching strongly-interacting large $r_s$ regime at higher carrier densities.} 
Crucially, the elevated electron densities imply that the electronic state is less susceptible to effects from sample disorder 
(to be discussed in more detail below). This situation may be compared with the 2DEGs historically realized in semiconductor quantum wells, which need much lower densities to access the WC regime and correspondingly require extremely clean samples (see Table \ref{tab:params}). 

%Experimentally, 2DEGs have long been studied in semiconductor heterostructures and quantum wells. Owing to their material parameters -- specifically the effective masses and dielectric constants; see Table~\ref{tab:params} -- these systems must be extremely clean to access the large $r_s$ regime required to observe WCs without being inhibited by sample disorder. In contrast, monolayer TMDs, due to their relatively large effective masses and reduced dielectric screening, reach the strongly-interacting large $r_s$ regime at much higher electron densities. While disorder remains an important issue (to be discussed in more detail below), the increased electron densities in TMDs imply that the electronic state is less susceptible to strong disorder effects. 

In addition to favorable band structures and material parameters, the high degree of tunability of modern 2D materials offers routes by which to enhance the stability of electron crystals, as well as opening the door to engineering more structured crystals exhibiting novel behaviors, beyond those of a single monolayer. The most direct example of this is stacking 2D materials to form multilayer structures, where the interalyer distances, tunneling, and twist angle between layers can all be controlled experimentally. The simplest case is a (non-twisted) bilayer of two Coulomb-coupled 2DEGs. By varying the ratio of the interlayer spacing $d$ to the interparticle spacing $a$, new WC geometries can be realized \cite{zhou2021,goldoni1996,rapisarda1996} (see also Sec.~\ref{sec:wc_outlook}), and the stability of the crystal is enhanced owing to commensurate locking of the two layers \cite{swierkowski1991,zhou2021,goldoni1996} (Fig.~\ref{fig:Fig1}{\bf e}). Utilizing optical probes, the emergence of bilayer WCs has been confirmed in bilayer MoSe$_2$ \cite{zhou2021}.
%Indeed, such bilayer systems have been demonstrated to host bilayer WCs, as revealed via optical methods in bilayer MoSe$_2$ \cite{zhou2021}. 

%For instance, electrons crystallize into a square lattice geometry when $d/a \sim$ 1. Such bilayer crystal remain stable to higher electron densities \cite{swierkowski1991,zhou2021,goldoni1996}, offering a significant advantage. These new geometries are accompanied by novel magnetic ground states --  also tunable via $d/a$ –- that are not realizable in an isolated layer. Importantly, the magnetic phases in the bilayer WC are associated with higher energy scales than those in the monolayer, as the bilayer crystal remains stable to higher electron densities \cite{swierkowski1991,zhou2021,goldoni1996}, offering a significant advantage.

When the 2D layers are twisted relative to each other, 
%Another unique capability of 2D layered systems is the ability to introduce a twist angle between layers.  
the resulting moiré patterns %in such twisted multilayers 
create a tunable external periodic potential experienced by the electrons (a similar effect can also arise in lattice-mismatched layers, or lattice mismatch with a substrate). The geometry, strength, and filling (the number of electrons per moiré unit cell) of the external potential serve as control parameters that can be used to stabilize the formation of electronic crystals and influence their properties. In the case of TMD heterobilayers, several groups have observed the formation of electronic crystals at certain rational fillings of the moiré superlattice in twisted WSe$_2$/WS$_2$ \cite{regan2020,xu2020,huang2021,jin2021,li2021,li2021b,li2024,shabani2021}. Similar observations have recently been made in pentalayer graphene \cite{walters2025,su2025}.  These moiré electron crystals -- or ``generalized Wigner crystals'' \cite{hubbard1978} -- exhibit properties distinct from those formed in the absence of the moiré potential. For instance, the presence of the potential leads to the formation of new types of crystals beyond the simple triangular lattice, such as stripe \cite{jin2021,li2021} and honeycomb \cite{li2021} lattices at 1/2 and 2/3 filling of a triangular moiré lattice, respectively. %The presence of the moiré potential is also expected to significantly modify the magnetic properties of the electron crystals \cite{motruk2023,yang2024_MI,yang2024_Honeycomb,biborski2025,esterlis2025b}. Interpolating between the limits of weak and strong moiré potential may shed light on the evolution of the electronic system from the crystal to the ``Mott-insulator” regime.

Yet another interesting feature of electron crystallization in 2D materials is the interplay of strong electronic correlations with topology of the host electronic band. Recent experiments on both crystalline and moiré graphene multilayers \cite{lu2024,lu2025,walters2025,su2025} have provided evidence for the formation of topological electron crystals at zero magnetic field, where the crystallization appears to coexists with a quantized Hall conductance. Intriguingly, the presence/absence of the quantized Hall effect has been shown to be tunable by external displacement and magnetic fields. 

\subsection{Real materials: deviations from the ideal 2DEG}

\begin{figure*}[t]
	\includegraphics[width=0.99\textwidth]{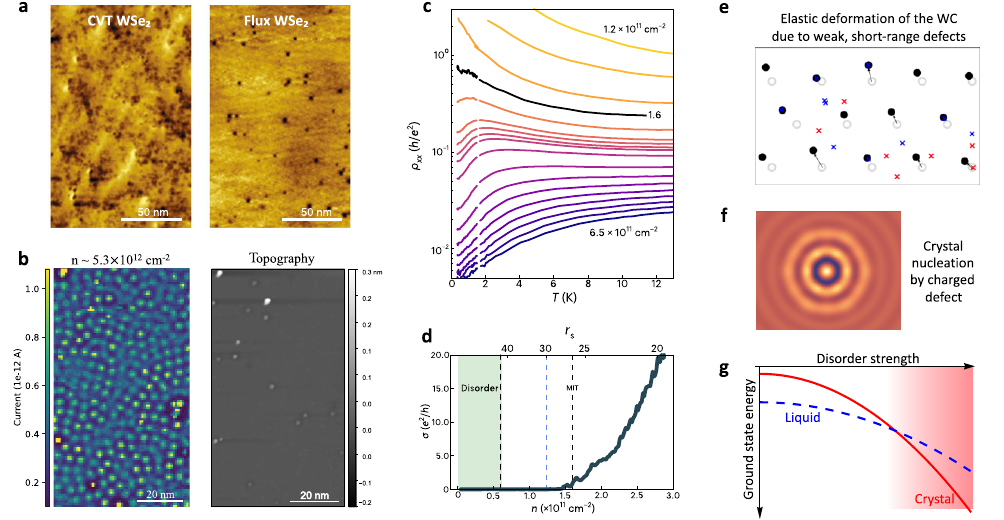}
	\caption{{\bf Influence of disorder on crystalline electronic phases in 2D materials.} ({\bf a}) Topographic images of CVT-grown (left) and flux-grown (right) WSe$_2$ crystals, visualizing appreciable difference in disorder density. ({\bf b}) STM image of an electronic Wigner crystal in hole-doped MoSe$_2$ bilayer (left) and a topography image of the same, undoped MoSe$_2$ flake (right), enabling direct visualization of defect-induced deformations of electronic lattice. ({\bf c}) Temperature- and electron-density dependent conductivity of high-mobility electron-doped WSe$_2$ monolayer, showing clear signature of metal-to-insulator transition ({\bf d}). ({\bf e}) Elastic deformation of a WC in the presence of weak, short-range attractive (blue) and repuslive (red) impurities. ({\bf f}) Schematic showing nucleation of a local Wigner ``crystallite" by a charged impurity in the liquid phase of the 2DEG; the expected $2k_F$ Friedel oscillations set in far from the impurity. ({\bf g}) Energy competition between the homogeneous electron liquid and crystal; weak disorder may tend to locally stabilize the crystal over the liquid phase owing to the gain in elastic energy. Panels {\bf a}, {\bf b}, {\bf c}--{\bf d} adapted from Refs.~\cite{rhodes2019},~\cite{xiang2025},~\cite{Pack2024}.}  \label{fig:Fig3}
\end{figure*}

%\textcolor{blue}{Some figure illustrating the disorder effects on the WC formation would be useful. One idea is to use some STM-visualization data from Feng Wang's paper.}

%Realistic perturbations in real materials: disorder, gates, ...

%\begin{itemize}
%\item Role of the disorder: in real, disordered-systems, we are somewhere between the true Wigner crystalization and Andreason localization; we shouuld comment on how this affects the stability of the crystaline phases probed in the experiment (critical $r_s$, critical temperature etc.)
%\item Role of number of layers in the Wigner crystal formation: by redistributing electrons between different layers, it is energetically more favorable for them to form the crystal.
%\end{itemize}
In real materials, the intricate interplay between electron-electron interactions and disorder plays a critical role in determining the properties of the WC, as well as the electron liquid to WC transition. Sufficiently strong disorder will induce Anderson localization, where electron wavefunctions become localized due to interference from fluctuations in the disorder potential. Although Anderson localization also creates an incompressible state, it differs fundamentally from a WC, which is characterized by strong positional correlations and a periodic electron arrangements. 

Experimentally, the amount of disorder in actual TMD devices depends on various factors, including synthesis methods, growth conditions, dielectric environment, and the fabrication of vdW heterostructures. For example, in TMDs prepared via mechanical exfoliation or chemical vapor deposition (CVD), defect densities can range from $10^{12}$ to above $10^{13}$ cm$^{-2}$ \cite{rhodes2019}. In comparison, high-quality graphene samples can have disorder densities as low as $10^{9}$ cm$^{-2}$ \cite{decker2011,xue2011, feldman2012}. The flux growth method has shown promise in reducing both charged and isovalent defect densities in TMD samples to below $10^{10}$ and $10^{11}$ cm$^{-2}$, respectively \cite{liu2023two,cevallos2019} (see Fig.~\ref{fig:Fig3}{\bf a}). This suggests that such samples could exhibit charge disorder well below the critical densities for WC formation, making them particularly promising for studying WCs in the clean limit. In addition to minimizing intrinsic disorder in the TMDs themselves, controlling the surrounding environment is crucial. For instance, encapsulating TMDs in hexagonal boron nitride (hBN) or suspending them has been shown to effectively suppress extrinsic disorder \cite{dean2010}.

%\textcolor{red}{Discuss the role of disorder in the melting temperature enhancement observed in 2D (unlike GaAs).}

While strong disorder leads to Anderson localization, a small amount of disorder can (locally) enhance the stability of the WC due to the energy gain from impurity pinning,  lowering the critical $r_s$ and increasing its melting temperature \cite{chui1995,vu2022,ahn2023}. Pictorially, one may imagine a pristine WC that becomes a ``glassy" localized state pinned by the disorder, while retaining short-range order \cite{chakravarty1999, voelker2001}. Electron localization is reinforced by the combined effects of Coulomb interactions and impurity potentials, effectively stabilizing the WC phase. See Fig.~\ref{fig:Fig3}{\bf e}-{\bf g}. The properties of such pinned WCs are well-studied \cite{ruzin1992,chitra1998,*chitra2001,*chitra2005}, and the additional pinning resonance has been a hallmark signature of the WC phase in optical absorption measurements on GaAs-based 2DEGs \cite{Williams1991,li1997,li2000,Ye2002,Chen2006,knighton2018}. Experimentally, a decreased critical $r_s$ and elevated critical temperatures have been found in MoSe$_2$ monolayers \cite{smolenski2021,sung2025} and bilayers \cite{xiang2025}, suggesting an important role of the disorder. 

Intriguingly, recent scanning tunneling microscopy experiments have directly imaged the disordered WC and the melting transition \cite{xiang2025}, allowing detailed investigations of the interplay between disorder and correlations. In this context, a particularly interesting direction concerns the connection between the WC transition and well-documented anomalies in the transport of strongly correlated 2DEGs \cite{spivak2010}, most notable of which is the metal-insulator transition. Local imaging, combined with recent progress in contacting TMD monolayers for transport studies (see Fig.~\ref{fig:Fig3}{\bf c}-{\bf d}) promises to shed light on these fundamental questions. 

In practice, the WCs realized in present 2D materials likely do not fit neatly into either the categories of ``strong" or ``weak disorder" described above \cite{huang2024}, and careful analysis,  along with reasonable quantitative criteria, is thus required to distinguish between the WC and Anderson (or Efros-Shklovskii \cite{efros1975}) insulator phases. This is especially true when considering more subtle phenomena such as the nature of the transition between the liquid and WC phases and possible two-phase coexistence \cite{joy2025}. We also note that the availability of \textit{local} probes for studying 2D materials (as elaborated upon below) means that disorder does not immediately preclude the possibility of studying WC physics.

Beyond quenched disorder, there are other perturbations to the idealized clean 2DEG that are especially relevant to the WCs realized in layered 2D materials. These include screening of the Coulomb interaction by nearby gate electrodes -- which render the Coulomb interactions short-range at separations larger than the distance to the gate and increase the critical $r_s$ \cite{valenti2025} -- and electron-atomic phonon coupling, which has recently been argued to be more important for the properties in WCs realized in 2D materials than in conventional quantum well systems \cite{tan2024}. 

%\textcolor{red}{Highlighting the role of local probes and strong correlations as an advantage of 2D platform vs old systems}

%\textcolor{red}{The following paragraph should be moved down, preferably to the outline.}

%Bilayer WCs feature a richer phase diagram compared to their monolayer counterparts and have attracted intense theoretical interest. However, experimentally creating two electrostatically coupled electron layers with precise control over their separation and doping densities has proven challenging in conventional materials. The advent of 2D materials and vdW heterostructures have opened exciting avenues to overcome these challenges. In recent experiments, two monolayers of MoSe2 are separated by a thin layer of hBN ($\approx$1nm) and then encapsulated in a dual-gate geometry, enabling independent control of electron densities in each layer. Intriguingly, by characterizing their electronic compressibility via optical spectroscopy, insulating states have been observed at symmetric (1:1) and asymmetric (4:1 and 7:1) electron doping ratios between the two MoSe2 layers at cryogenic temperatures. These distinct density ratios correspond to bilayer WC composed of two interlocking commensurate triangular electron lattices stabilized by interlayer interactions. Remarkably, these bilayer Wigner crystal phases exhibit enhanced stability compared with their monolayer counterpart, undergoing both quantum and thermal melting transitions at electron densities as high as $~10^{12}$ cm$^{-2}$ and temperatures reaching almost 40 K.

\section{Sensing methods}
\label{sec:wc_sensing}
%\item[(2)] \textbf{Sensing methods}
%\textcolor{blue}{
%\begin{itemize}
%\item We can start with conventional transport, which can be potentially useful, but is not allowing for proving the charge-order.
%\item We them move to optics, emphaizing different approaches. We can start with methods for detecting electronic compressibility (which serve similar purpose as transport experiments, but is more local). Then we move to umklapp spectroscopy, allowing for detecting long-range charger-order.
%\item Here we should mention about local scanning probe sensing, such as STM or single-electron transistors
%\end{itemize}}

The hallmark of electronic crystals is the presence of long-range charge order. Even though this order is conceptually identical to that of atoms in regular crystals, the electronic ones are notoriously difficult to probe. This is primarily due to their much lower densities, which renders regular crystallography techniques, such as an X-ray diffraction, not efficient when it comes to detection of electronic crystals. Early experimental efforts thus relied on indirect signatures of WCs: instead of probing the charge order, they focused on the compressibility. As discussed in the previous section, the WCs in real materials are incompressible due to pinning of the electric lattice by disorder. For this reason, the WC acts as an insulator, and its formation thus manifests as an increased resistivity in DC or low-frequency transport experiments~\cite{Andrei1988,Goldman1990,Yoon1999}. Another consequence of disorder pinning is the emergence a pinning-mode resonance in high-frequency AC transport experiments, which corresponds to collective oscillations of an electronic crystal about the disorder potential minima ~\cite{Williams1991,li1997,li2000,Ye2002,Chen2006,knighton2018}. 

%While ideal WCs are predicted to be compressible~\cite{xxx}, in real materials the electronic crystals are always pinned to the underlying lattice by disorder. As a result, the WC is incompressible, and hence acts as an insulator; its formation thus manifests as an increased resistivity in DC or low-frequency transport experiments. This method has been employed in numerous experiments for detecting the WC at both zero magnetic field~\cite{xxx}, and at finite $B$-fields, where the competition between charge ordering and the formation of quantum Hall phases gives rise to oscillations in longitudinal resistivity that rises or vanishes for the two phases~\cite{xxx}. Another consequence of disorder pinning is the emergence a pinning-mode resonance --- typically in the microwave frequency range --- which corresponds to collective oscillations of an electronic crystal about the disorder potential minimum. Such a resonance disappears upon melting, often serving as a fingerprint of the WC formation in high-frequency AC transport experiments.

While these techniques have served as workhorses for WC explorations in conventional materials (e.g., GaAs), their application in the context of vdW heterostructures turned out to be more challenging. First, due to limited lateral extension and spatial inhomogeneities originating from mechanical stacking process, standard transport experiments often prove unsuccessful, as they inherently average electronic properties over the entire sample. Owing to difficulties in making high-quality electrical contacts to semiconducting 2D materials, such measurements of low-density electronic crystals have been mostly limited to graphene-based systems, although there are has been recent progress in this direction~\cite{Park2023,Pack2024,Kang2024}, including very recent THz spectroscopy experiments of the AC conductivity of a WC in a TMD monolayer~\cite{chen2025}. Nevertheless, from this perspective, local sensing techniques offering in-situ selection of investigated spatial areas are clearly advantageous. This includes both optical and scanning probe methods that are discussed below. %{\color{red} IE: Highlight that these local methods are perhaps even better for WC detection, since probes like transport average over the whole system.}

\subsection{Optical spectroscopy of crystalline electronic phases in 2D materials}
With its sub-micron spatial resolution, confocal spectroscopy provides a unique compromise between experimental complexities and local access to electronic phases. TMD-based systems are particularly well-suited for this approach thanks their strong exciton binding energies, which are enhanced with respect to conventional materials for exactly the same reasons that are responsible for more prominent electronic correlations: weak dielectric screening and relatively large carrier effective masses. This renders the excitons in these materials as robust bosonic impurities for sensing electronic phases even at relatively large carrier densities $\sim10^{12}\ \mathrm{cm}^{-2}$~\cite{Sidler2017,Efimkin2017}. In the presence of charge carriers, the excitons form trions that give rise to repulsive and attractive Fermi polarons (AP), which dominate the luminescence and absorption spectra.

\begin{figure*}[t]
	\includegraphics[width=0.99\textwidth]{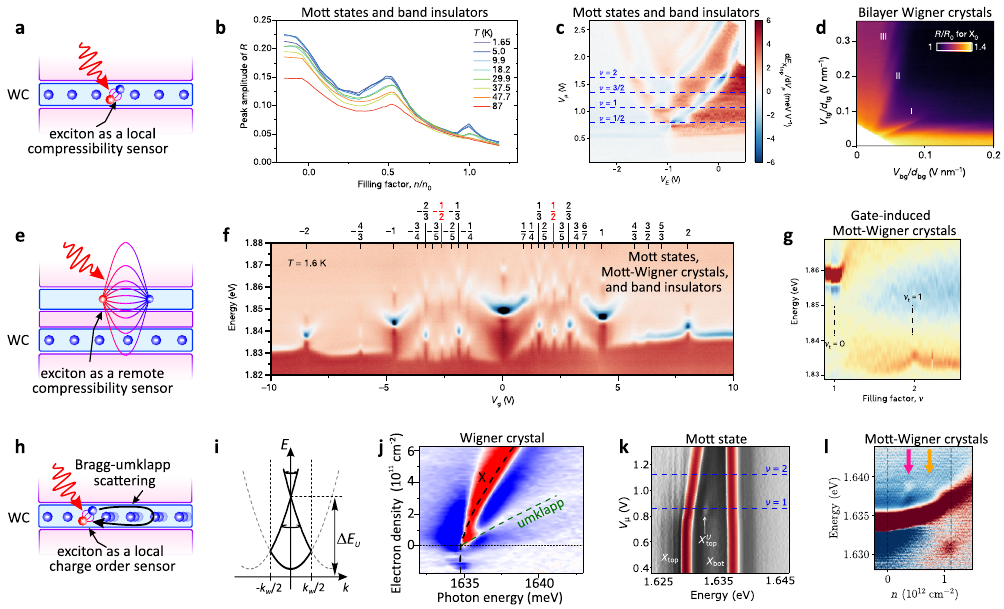}
	\caption{{\bf Optical sensing of charge-ordered electronic phases.} ({\bf a})~Local compressibility sensing with excitons residing in the same layer as the electron system. Formation of incompressible phases such as Mott states ({\bf b,c}) or WCs ({\bf d}) gives rise to changes in the amplitude or linewidth of the excitonic resonance. ({\bf e})~Remote optical sensing of electronic compressibility with Rydberg excitons in an adjacent layer. Changes in the compressibility related to the formation of Mott or Mott-Wigner phases ({\bf f,g}) result in drastic changes in spectral positions and weights of Rydberg excitons. ({\bf h})~Bragg-umklapp scattering of excitons off the crystalline electronic phase, allowing for optical sensing of long-range charge order. The interaction between excitons and ordered electronic lattice folds back excitonic bands, giving rise to new umklapp optical resonances, whose energy offset from the main exciton transition is directly determined by the lattice constant of Wigner ({\bf j}), Mott ({\bf k}), or Mott-Wigner ({\bf l}) states. Panels {\bf b}, {\bf c}, {\bf d}, {\bf f}, {\bf g}, {\bf i}, {\bf j}, {\bf k}, {\bf l} adapted from Refs.~\cite{tang2020},~\cite{shimazaki2020},~\cite{zhou2021},~\cite{xu2020},~\cite{gu2024},~\cite{smolenski2021},~\cite{shimazaki2021},~\cite{kiper2025}.} \label{fig:Fig4}
\end{figure*}

The most straightforward application of these optical excitations is local compressibility sensing (Fig.~\ref{fig:Fig4}{\bf a}). Experimentally, whenever the electron system in a TMD-based system becomes incompressible, the optical resonances display a sharp change in their intensity, energy, or linewidth. This applies not only to charge-ordered phases (Fig.~\ref{fig:Fig4}{\bf b}-{\bf d}) such as WCs~\cite{zhou2021}, Mott states~\cite{tang2020}, or Mott-Wigner crystals~\cite{Campbell2024}, but also to other gapped states, such as integer quantum Hall liquids~\cite{Smolenski2019,Liu2020,Li2020}. Excitonic sensing can be also operated non-locally to optically detect electrical conductivity~\cite{regan2020}. Moreover, it remains efficient in the case of multilayer electronic phases, even if they are hosted by layers of the same material separated by insulating spacers. This is due to natural strain variations in vdW heterostructures, which typically result in differences between energies of excitons in various layers, thus enabling layer-selective spectroscopic readout of electronic compressibility using the excitonic transition originating from a specific layer. Such a method has recently allowed for probing the formation of bilayer Mott states in the system of MoSe$_2$ monolayers with a monolayer hBN spacer~\cite{shimazaki2020} (Fig.~\ref{fig:Fig4}{\bf c}). 

Another unique aspect of optical experiments is that they provide a direct interface to the spin state of correlated electrons. Owing to spin-valley locking in TMD-based systems, the circular polarization of optical excitations is determined by the spin polarization of resident electrons~\cite{back2017}. This enables direct optical sensing of electronic magnetism, which has recently been exploited to probe the evolution of magnetic susceptibility across the solid-to-liquid electron phase transition in a MoSe$_2$ monolayer~\cite{sung2025}, shedding light on the evolution of exchange interactions across the WC melting transition. A similar technique has also been employed to study the collective magnetism of charge-ordered electronic phases in moir\'e bilayers~\cite{xu2020,ciorciaro2023,tang2023,tao2024}.

%In the latter, moir\'eless system, bilayer WCs have been observed not only at symmetric (1:1), but also asymmetric (4:1 and 7:1) electron doping ratios between the two MoSe2 layers seperated by a thin hBN layer, thus demonstrating the importance of interlayer interactions stabilizing interlocked commensurate triangular electron lattices.

In addition to the above optical sensing schemes involving electrons and excitons from the same layer, the electronic correlations in a given layer can be also probed using excitons residing in an adjacent layer. For example, by probing the onset of doping in this sensor layer, it is possible to optically detect the chemical potential of a remote electron system~\cite{Xia2024}. When the sensor layer remains charge-neutral, its Rydberg excitonic states can serve as sensitive optical sensors of incompressible electronic phases (Fig.~\ref{fig:Fig4}{\bf e}) not only in proximal TMD bilayers systems~\cite{xu2020,gu2024} (Figs.~\ref{fig:Fig4}{\bf f,g}), but also in layered materials that are otherwise optically-inaccessible, such as graphene~\cite{hu2023,he2024}.

%particularly useful are excited Rydberg excitonic states featuring large Bohr radii, whose emission energies and oscillator strengths are strongly dependent on the ability of proximal electrons to screen the Coulomb interactions~\cite{...}. This renders the Rydberg states as sensitive optical sensors of incompressible electronic phases not only in TMD bilayers systems, but also in layered materials that are otherwise optically-inaccessible, such as graphene.

Despite their versatility, the above methods cannot be employed for detecting a crystalline electronic lattice directly. Much more powerful in this regard is Bragg-umklapp spectroscopy. At the heart of this approach is the periodic potential experienced by the exciton interacting with an electronic crystal (Fig.~\ref{fig:Fig4}{\bf h}). This allows for an optically-inactive exciton of finite momentum to get folded back to the light cone, provided that its momentum matches the reciprocal WC lattice vector $k_W$ (Fig.~\ref{fig:Fig4}{\bf i}). This gives rise to a new \textit{umklapp} resonance in the absorption spectrum, which is blueshifted by the kinetic energy of Bragg-scattered excitons $\hbar^2 k_W^2/2m_X^*$. This resonance not only serves as a direct signature of a long-range crystalline order, but also as a quantitative probe of the underlying lattice constant determining $k_W$. This method has been exploited both for sensing charge-ordered phases, including WCs in translationally-invariant mononolayer~\cite{smolenski2021} (Fig.~\ref{fig:Fig4}{\bf j}) and bilayer MoSe$_2$ systems~\cite{zhou2021}, as well as Mott and Mott-Wigner states in TMD bilayers hosting moir\'e potentials~\cite{shimazaki2021,kiper2025} (Figs.~\ref{fig:Fig4}{\bf k,l}). In the latter case, the key prerequisite for applicability of the method is the lack of periodic potential for excitons in the absence of electrons, which is realized in systems where conduction and valence bands experience moir\'e potentials of identical periodicity, such as twisted TMD bilayers with a monolayer hBN spacer~\cite{shimazaki2020,shimazaki2021} or TMD monolayers interfaced with twisted hBN bilayer~\cite{kiper2025}. 

\subsection{Local probe sensing of crystalline electronic phases in 2D materials}

To access WCs on length scales shorter than the optical wavelength, scanning probe sensing becomes the technique of choice. In this method, a sharp tip is brought in close proximity to a device, thereby not only achieving outstanding spatial resolution but also allowing to  probe electronic properties without the need for high-quality electrical contacts. Among the multitude of local probe schemes (see Ref.~\cite{nuckolls2024} for a comprehensive recent review on their usage in 2D materials), those applicable for the spectroscopy of charge-ordered phases can be classified into two main categories depending on the physical observable they give access to. 

\begin{figure}[t]
	\includegraphics[width=0.49\textwidth]{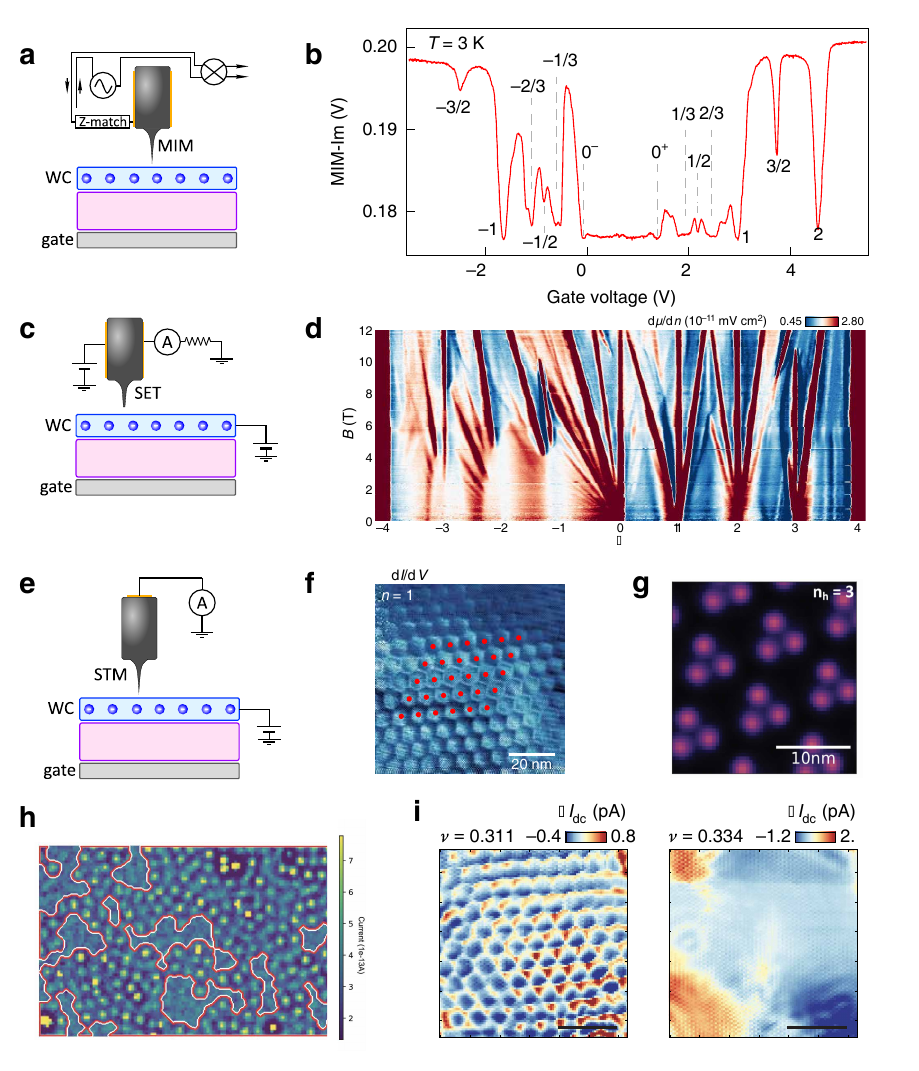}
	\caption{{\bf Local probes of charge-ordered electronic phases.} ({\bf a}) MIM and ({\bf c}) SET sensing of electronic compressibility, enabling detection of incompressible electronic phases in WSe$_2$/WS$_2$ bilayers ({\bf b}) or graphene-based systems ({\bf d}) with $\sim$100~nm spatial resolution. ({\bf e}) STM probing of charge ordering with spatial resolution of single nanometers. This method enabled visualization of real space electronic lattices of Mott states in WSe$_2$/WS$_2$ bilayers ({\bf f}), Wigner molecules in hole-doped twisted WS$_2$ bilayers ({\bf g}), WCs in hole-doped natural MoSe$_2$ bilayers ({\bf h}), and electronic crystals in graphene subjected to strong magnetic fields ({\bf i}). Panels {\bf b}, {\bf d}, {\bf f}, {\bf g}, {\bf h}, {\bf i} adapted from Refs.~\cite{huang2021},~\cite{xie2021},~\cite{li2021},~\cite{li2024b},~\cite{xiang2025},~\cite{tsui2024}.} \label{fig:Fig5}
\end{figure}

The first one includes techniques such as microwave impedance microscopy (MIM) (Fig.~\ref{fig:Fig5}{\bf a}) or scanning electron transistor (SET) spectroscopy (Fig.~\ref{fig:Fig5}{\bf c}) that enable local measurements of macroscopic quantities like electronic conductivity or compressibility on the length scales $\sim100$~nm. These techniques, albeit not being able to reveal the periodic electronic order, allow for distinguishing insulating and metallic phases, and have been successfully employed to sense the formation of Mott states, Mott-Wigner crystals~\cite{huang2021} (Fig.~\ref{fig:Fig5}{\bf b}) or $B$-field-induced Wigner crystals~\cite{xie2021,ji2024} (Fig.~\ref{fig:Fig5}{\bf d}) in various vdW structures.

The second, more direct scheme, involves scanning tunneling microscopy (STM) (Fig.~\ref{fig:Fig5}{\bf e}) that offers spatial resolution at the level of single nanometers and uniquely enables visualization of the real space structure of electronic crystals. In one of the first STM applications to electron crystals, the moir\'e bilayer hosting a generalized WC was not exposed directly to the STM tip, but covered with a thin insulating spacer and a doped graphene layer~\cite{li2021}. In this configuration, the tunneling current between the tip and graphene layer is spatially modulated due to interlayer interactions with the proximal WC, enabling direct mapping of the corresponding charge distribution with a spatial resolution limited by the hBN spacer thickness~\cite{li2021} (Fig.~\ref{fig:Fig5}{\bf f}). More recently, this scheme has been further improved to enable direct sensing of WCs without the need for an intermediate metallic layer: by carefully balancing the band edge and vacuum energy levels in the tip and the probed layer, the tip-induced perturbation of the electron system was reduced to an extent that allowed for non-invasive sensing. This has enabled imaging the formation of Wigner molecules at high filling factors of a deep moir\'e potential~\cite{li2024b} (Fig.~\ref{fig:Fig5}{\bf g}),  quantum melting of the hole WC in bilayer MoSe$_2$~\cite{xiang2025} (Fig.~\ref{fig:Fig5}{\bf h}), and the competition between the WC and fractional quantum Hall states in graphene subject to a Landau-quantizing magnetic field~\cite{tsui2024} (Fig.~\ref{fig:Fig5}{\bf i}).

\section{Outlook}
\label{sec:wc_outlook}

%{\color{blue} highlight enhanced stability and magnetism of exotic crystals. kinetic magnetism}

\subsection{More exotic electron crystals}

%\begin{itemize}
%\item Dirac systems: In graphene-based systems, the Dirac dispersion of electrons near the $K$ and $K'$ points of the Brillouin zone and the Berry curvature may also qualitatively modify the properties of the electron crystals that may form. A fundamental – and experimentally relevant – question in such system concerns the interplay between the strong electronic correlations of the electron crystal and the topology of the host electronic band. \cite{joy2022,joy2023_bernal,joy2025_chiral,kim2025}
%\item Multivalley systems: In multivalley systems with anistropic mass tensors, interesting form of pseudo-spin order  (associated with the valley degree of freedom) have been predicted to occur at elevated temperature scales. \cite{calvera2003}
%\end{itemize}

The versatility of layered vdW materials opens up entirely new avenues for studying electron crystallization. Even in the single layer limit, intriguing possibilities remain beyond those discussed above. For example, interband screening in gapped monolayer graphene can strongly influence electronic charge-ordering \cite{joy2022}, while multivalley systems with anisotropic effective mass tensors may host interesting forms of pseudo-spin order  (associated with the valley degree of freedom) at elevated temperature scales \cite{calvera2003}.

The modularity of 2D materials also offers unprecedented opportunities for heterostructure design, where stacking layers with distinct properties to form coupled multilayers significantly expands the accessible phase space. In addition to stabilizing novel phases, these structures can provide access to the fascinating phenomenology inherent to electronic crystals, such as their magnetic properties and excitation spectra. 

A particularly simple example of this, which has been mentioned already, is the Coulomb-coupled bilayer WC with like charges in both layers. Here two additional axes appear in the ground state phase diagram: the interlayer separation and the density imbalance between layers. For commensurate layer densities, interlayer interactions stabilize a variety of geometries, including square and rectangular lattices \cite{goldoni1996,zhou2021}, which may be directly imaged using scanning probe techniques. Additionally, gate tuning of individual layer concentrations allows for controlled doping away from commensurate densities. In this situation, a sufficiently small density imbalance has been theoretically predicted to result in defect-doped (interstitial or vacancy) \textit{metallic} electron crystal ground states \cite{zhuang2025}, which may themselves have further ordering instabilities.

 %Furthermore, bilayer Wigner crystals with a slight density imbalance are predicted to host electronic interstitials and/or vacancies even at zero temperature. It was suggested that these intrinsic point defects may induce a self-doping effect and facilitate kinetic magnetism before the quantum melting of WCs.

\begin{figure}[t]
	\includegraphics[width=0.49\textwidth]{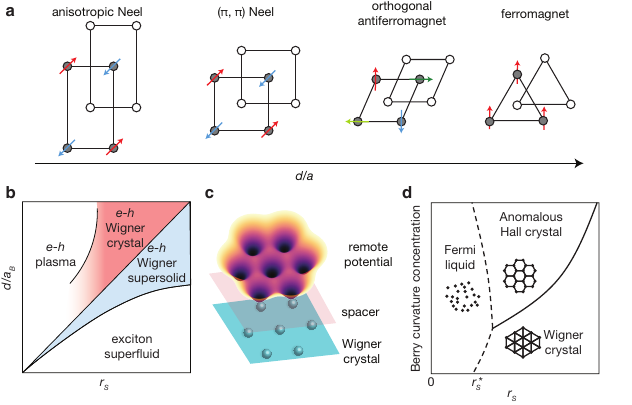}
	\caption{{\bf Outlook} ({\bf a}) Bilayer Wigner crystals are predicted to host a variety of structural and magnetic phases depending on the interlayer coupling. ({\bf b}) An electrostatically coupled electron–hole bilayer system can support exotic phases, including bilayer Wigner crystals, excitonic superfluids, and more exotic Wigner supersolids. ({\bf c}) A Wigner crystal can imprint a periodic electrostatic potential on another neighbouring layer as a source to create new ordered phases.({\bf d}) A schematic phase diagram as a function of interaction strength and Berry curvature concentration, showing regions of Fermi liquid, conventional Wigner crystal, and anomalous Hall crystal phases. Panels {\bf a, b, d} adapted from \cite{esterlis2025},\cite{joglekar2006},\cite{zheng2024sublattice}} \label{fig:Fig6}
\end{figure}

Probing and manipulating the magnetic (either spin or valley pseudo-spin) states of WCs is another promising direction. While theory predicts a ferromagnetic ground state as $r_s \to \infty$ in the monolayer WC, in the experimentally relevant range of densities, the combination of relatively weak exchange interactions and a high degree of frustration suppresses magnetic ordering temperatures (see Sec.~\ref{sec:wc_general}). This makes experimental investigation of magnetic properties challenging, despite significant interest. 
Bilayer WCs offer a promising platform to address these challenges, as their higher critical densities lead to stronger exchange interactions. Moreover, recent theoretical work suggests that varying the interlayer coupling in different bilayer lattice configurations can yield distinct magnetic ordering \cite{esterlis2025} -- including ferromagnetic and multi-sublattice antiferromagnetic states (Fig.~\ref{fig:Fig6}{\bf a}) -- opening new avenues for controllable magnetism. In twisted multilayers, moiré potentials significantly modify the magnetic properties of electron crystals \cite{motruk2023,yang2024_MI,yang2024_Honeycomb,biborski2025,esterlis2025b}. Beyond conventional exchange dynamics, quantum tunneling of WC defects has recently been predicted to lead to kinetic magnetism with significantly elevated energy scales \cite{kim2022,kim2024,ciorciaro2023,morera2023}.

Another intriguing possibility involves oppositely charged 2D layers (electrons and holes) brought into close proximity \cite{szymanski1994, moudgil2002}. At large layer separations and for sufficiently dilue densities, carriers form interlocking triangular WCs; as the separation decreases, stronger interlayer interactions lead to the formation of bound electron-hole pairs with phase coherence; i.e., superfluidity. This transition has been predicted to proceed via an intermediate bosonic supersolid -- a quantum phase that combines crystalline order with superfluidity \cite{joglekar2006,boening2011} (Fig.~\ref{fig:Fig6}{\bf b}). %Theoretical predictions point to these as intermediates between WCs and exciton superfluids \cite{joglekar2006,boening2011} (Fig.~\ref{fig:Fig6}{\bf b}).
How to precisely engineer the optimal interlayer and intralayer Coulomb interactions -- and how to detect these phases experimentally -- remains an open and exciting frontier.
 
WCs themselves can also act as sources of tunable periodic potentials for adjacent layers, similar to the remote imprinting of moiré potentials \cite{gu2024, kim2024_twistedhbn,kim2025moire,gu2025}. By dynamically tuning the WC, one can imprint a reconfigurable periodic landscape onto adjacent layers (Fig.~\ref{fig:Fig6}{\bf c}). This tunable potential enables the stabilization of a wide range of phases, including charge-ordered states, dipolar insulators \cite{chen2022excitonic,gu2022dipolar}, and states with fractal energy spectra such as Hofstadter butterflies \cite{zheng1995}.

%Another exciting, yet still uncharted territory concerns collective spin physics of periodically-ordered electronic crystals. In prototypical single-layer WCs, the resulting magnetic phases, albeit theoretically predicted~\cite{todo}, are very difficult to explore due to very small exchange interactions between tightly-localized electrons in a lattice of the order of XXX. However, collective spin phases are expected to be much more robust for multilayer WCs that can exist for sizably lower $r_s$~\cite{todo}. \textcolor{blue}{TODO: extend}

%\textcolor{blue}{Figure depicting the implementations we discuss.}
%\begin{itemize}
%\item Electronic crystallization in a system when one layer hosts a strong moire potential, and then other is pristine.
%\end{itemize}
Finally, the interplay between electron crystallization and the topology of the host electronic band presents a new direction for exploration. With regards to magnetism, recent semiclassical calculations have shown that Berry curvature of the host electronic band can modify the spin dynamics qualitatively, leading to chiral terms in the effective exchange Hamiltonian that may stabilize chiral spin-density wave or chiral spin liquid phases \cite{kim2025,joy2025_chiral}.
Perhaps even more dramatic, the breaking of translation symmetry in a band with Berry curvature via electron crystallization can lead to back folded bands with non-zero Chern numbers and corresponding quantized Hall responses. While related ``Hall crystal" phases were proposed decades ago in quantum Hall systems \cite{tesanovic1989}, recent experiments in both graphene/hBN moiré superlattices \cite{lu2024,lu2025,walters2025} and twisted bilayer-trilayer graphene \cite{su2025} have observed a quantized Hall resistance and vanishing longitudinal resistance at \textit{zero magnetic field} that appears to coexist with some form of electron crystallization. In these systems, a quantized Hall effect appears at both commensurate fillings of the moiré lattice \cite{lu2024, su2025, walters2025} and,  more surprisingly, over extended \textit{ranges} of filling \cite{lu2025}, suggestive of a continuous translation symmetry breaking. The temperature dependence of the resistances and the threshold current-voltage characteristics also resemble those of WCs. These ``anomalous Hall crystals" (AHCs), featuring both charge order and nontrivial topology in the form of a quantized Hall response, are now the subject of active theoretical investigation \cite{zheng2024sublattice,zhou2024,tan2024_ahc,dong2024,soejima2024,dong2024stability,tan2025,sheng2024quantum,zhou2025,soejima2025jellium,desrochers2025} (Fig.~\ref{fig:Fig6}{\bf d}). One could also imagine stabilizing an AHC by placing it adjacent to a topologically trivial WC that serves as a tunable ``pinning potential", as described in the preceding paragraph.

\subsection{How it can be observed}

New experimental approaches to studying WCs promise significant advances in understanding and controlling these intriguing phases. Although transport studies have been challenging, future improvements in metal-2D semiconductor contacts \cite{wang2022making} could lead to a detailed investigation of the thermodynamic stability and nonlinear transport properties of WCs. Microwave spectroscopy techniques \cite{chen2003, brem2022}, which can characterize the vibrations of WCs, including their phonons and pinning/de-pinning dynamics, could provide critical information on their dynamic properties and stability.

Scanning probe measurements have proven powerful for studying WCs and can be further enhanced by incorporating additional capabilities. For instance, integrating optical or microwave methods with scanning probe microscopy \cite{hillenbrand2025,cocker2016} not only enables nanoscale mapping of WC's response at such wavelengths, but also provides a powerful approach to manipulate electrons within WCs through optical or spin resonances. Furthermore, the recent development of \cite{inbar2023}, where a twisted 2D junction serves as the tunneling junction in scanning tunneling microscopy (STM), may reveal critical insights into the properties of WCs in the momentum space.

%%Given that the spin polarization degree of the electrons in TMD systems is encoded in the circular polarization degree of exciton-polaron optical transitions. 

Optical methods could also enable detection of collective spin physics. Owing to the locking between electronic spin orientation and helicity of excitonic resonances, by probing temporal correlations of photons interacting with a TMD layer, it might be possible to infer the dynamics of spin fluctuations of the WC electrons and study their evolution across the quantum phase transition to the liquid phase. This could reveal the formation of spin-ordered WC at sufficiently low temperatures, either with ferro- or anti-ferromagnetic spin arrangement. In the latter case, umklapp spectroscopy can uniquely enable one to probe the geometry and size of various spin sublattices, which will interact differently with excitons of a given circular polarization. This is expected to give rise to a fine structure of the umklapp resonance~\cite{salvador2022}.

Additionally, ultrafast optical techniques offer the capability to dynamically manipulate and control WCs. Such ultrafast control enables the exploration of transient phenomena and non-equilibrium states, potentially leading to novel quantum phases and transitions that are inaccessible under equilibrium conditions \cite{kim2020optical,sarkar2025}.

Collectively, these advanced methodologies could significantly deepen our understanding of both the ground and excited states of electronic crystals, and also open exciting new avenues for manipulating correlated electron systems more broadly.

\begin{acknowledgments}
IE was supported by the National Science Foundation (NSF) through the University of Wisconsin Materials Research Science and Engineering Center Grant No. DMR-2309000. YZ was supported by the U.S. Department of Energy, Office of Science, Office of Basic Energy Sciences Early Career Research Program under Award No. DE-SC-0022885 and the National Science Foundation under Award No. DMR-2145712.
\end{acknowledgments}

\bibliography{refs}

\end{document}